\documentclass{IEEEtran}
\usepackage{cite}
\usepackage{amsmath,amssymb,amsfonts}
\usepackage{algorithmic}
\usepackage{graphicx}
\usepackage{textcomp}
\usepackage{lipsum,mathtools,cuted}
\usepackage{xcolor}

\usepackage{float}
\usepackage{epsfig}
\usepackage{epstopdf}
\usepackage{subfigure,color}
\usepackage{dcolumn}
\usepackage{bm}
\usepackage{lineno}
\usepackage{accents}
\usepackage{multirow}
\usepackage[usestackEOL]{stackengine}
\usepackage{tabularx}
\usepackage{makecell}
\usepackage{dcolumn}% Align table columns on decimal point
\usepackage{bm}% bold math
\usepackage{booktabs}
\def\BibTeX{{\rm B\kern-.05em{\sc i\kern-.025em b}\kern-.08em
    T\kern-.1667em\lower.7ex\hbox{E}\kern-.125emX}}
    
\def\e{\begin{equation}}
\def\f{\end{equation}}
\def\_#1{{\bf #1}}

\def\.{\cdot}

\def\=#1{\overline{\overline #1}}
\def\Re{{\rm Re\mit}}

\def\@#1{_{\rm #1}}

\begin{document}
\title{Macroscopic Modeling of Anomalously Reflecting Metasurfaces: Angular Response and Far-Field Scattering}

%Far-field scattering from metasurface anomalous reflectors under arbitrary plane-wave illumination

%On the Macroscopic Modelling of Metasurfaces for Intelligent Radio Environments: Angular Response of Anomalous Reflectors

\author{A.~D{\'i}az-Rubio,~\IEEEmembership{Member, IEEE} and S.A.~Tretyakov,~\IEEEmembership{Fellow, IEEE}

\thanks{(Corresponding author: A.~D{\'i}az-Rubio)  This work was supported in part by
the European Commission through the H2020 ARIADNE project under grant
871464.}
\thanks{A.~D{\'i}az-Rubio and S.A.~Tretyakov 
are with the Department of Electronics and Nanoengineering, Aalto University, 
FI-00079 Aalto, Finland 
(e-mail: ana.diazrubio@aalto.fi; sergei.tretyakov@aalto.fi).}
}

\maketitle

\begin{abstract}
In view of extremely challenging requirements on the design and optimization of future mobile communication systems,  researchers are considering the possibilities of creation intelligent radio environments by using reconfigurable and smart metasurfaces integrated into walls, ceilings, or facades. In this novel communication paradigm, tunable metasurfaces redirect incident waves into the desired directions. To design and characterize such smart radio environments in any realistic scenario, it is necessary to know how these metasurfaces behave when illuminated from other directions and how scattering from finite-sized anomalous reflectors can be estimated. 
In this work, we analyze the far-field scattering of reflective metasurfaces and study the angular response of anomalous reflectors for arbitrary illumination angles. Using the surface-impedance model, we explain the dependence of the reflection coefficients of phase-gradient metasurfaces on the illumination angle and present numerical examples for typical structures. We also consider scattering from finite-size metasurfaces and define a route toward including the full-angle response of anomalous reflections into the ray-tracing models of the propagation channel. The developed models apply to other diffraction gratings of finite size.

\end{abstract}


\section{Introduction}
\IEEEPARstart{W}{ith} the development of mobile communications, there is a continuous challenge to improve the communications systems making them more efficient and versatile. 
New research directions point toward the use of intelligent radio environments supported by reconfigurable metasurfaces; see,  e.g., \cite{Liaskos,di2019smart,basar2019wireless,smart1,smart2,smart3,smart4,smart5,smart6}. 
In contrast to conventional wireless communication systems where only transmitters and receivers can be optimized, these novel approaches will potentially enable the controllability, programmability, and
optimization of the propagation channel.
Metasurfaces can control the properties of reflected and scattered fields and, by embedding tunable and active elements in the constitutive meta-atoms, allow changing their electromagnetic response, optimizing the functionality. 
Among different types of metasurfaces, reflective metasurfaces, typically backed by a metallic plane that blocks transmitted waves, are good candidates to be integrated into flat or curved surfaces of walls, ceilings, building facades, etc.

\begin{figure}[h]
\centerline{\includegraphics[width=0.9\columnwidth]{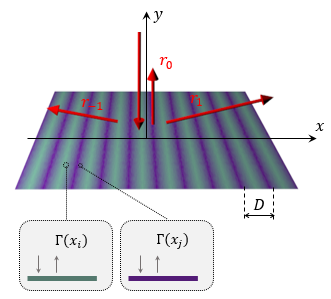}}
\caption{Schematic representation of the far-field scattering produced by a periodic reflective metasurface. In addition, a representation of the physical meaning of the local reflection coefficient is shown in the bottom.} 
\label{fig:Reflection}
\end{figure}

The operational principle of metasurfaces capable of reflecting incident waves into anomalous directions is similar to that of diffraction gratings. According to the Floquet theory, selecting the spatial periodicity of the structure, $D$, allows us to define the propagation directions of reflected modes, opening new possibilities for engineering reflection.  
If we illuminate an infinite periodic  metasurface  by a plane wave at the incident angle $\theta_{\rm i}$, the reflected field is  defined as a superposition of plane waves propagating in different directions according to the relation
\begin{equation}
\sin\theta_{{\rm r}n}=\sin\theta_{\rm i}+\frac{2\pi}{kD}n
%,\quad \cos\theta_{{\rm r}n}=\frac{\sqrt{k^2-(\sin\theta_{\rm i}+n2\pi/D)^2}}{k^2}
\label{trn}
\end{equation}
where $\theta_{{\rm r}n}$ are the reflection angles of the propagating harmonics of index $n=0,\pm 1,\pm 2,\dots$.
In the analysis of the scattering properties of metasurfaces, it is convenient to define the reflection coefficient of the metasurface as the ratio between the tangential components of the electric field of the incident plane wave and the reflected plane waves.
The existence of multiple directions of propagation does not allow using a unique reflection coefficient. For this reason,  we introduce a reflection coefficient for modeling metasurfaces as a combination of  the individual reflection coefficients for each propagating reflected mode:    
\begin{equation}
    R(\theta_{\rm i},x)=\sum_nr_n(\theta_{\rm i})e^{-jk(\sin\theta_{{\rm r}n}-\sin\theta_{\rm i})x}
    %e^{-jk(\cos\theta_{{\rm r}n}+\cos\theta_{\rm i})y} 
    \label{eq:macroscopic}
\end{equation}
%I suggest to define R at the surface only, this is how reflection coefficients are usually defined, and it is easier to understand.
where $r_n(\theta_{\rm i})=|r_n(\theta_{\rm i})|\exp{[j\phi_n(\theta_{\rm i})]}$ are the ratios of the complex amplitudes of the propagating Floquet harmonics of the reflected field and the tangential component of the incident electric field at the metasurface plane.  We call the coefficients $r_n$ the individual reflection coefficients for each propagating diffracted mode.

The effective reflection coefficient defined by Eq.~(\ref{eq:macroscopic}) describes the {\it macroscopic response} of the metasurface and it is useful for calculation of reflected and scattered fields in the far zone. However, it does not provide a direct route for designing the metasurface because the reflective properties depend also on the evanescent fields near the metasurface that are not included in this definition. The most common approach to design metasurfaces for manipulating the direction of reflected waves is to make the local reflection phase nonuniform along the metasurface. As it is known from the phased-array antenna theory, in the particular case of anomalous reflection (a plane wave is reflected breaking the reflection law, i.e., the reflection angle  $\theta_{\rm rd}$ is not equal to the incidence angle $\theta_{\rm id}$), the reflection phase should linearly vary compensating the phase mismatch between the incident and reflected waves.
The corresponding local reflection coefficient, defined as the ratio between tangential electric fields of the incident and reflected waves at each point of the metasurface, can be written as $\Gamma(x)=\exp[{j(\sin\theta_{\rm id}-\sin\theta_{\rm rd})kx]}$. Here, $k=\omega\sqrt{\mu\varepsilon}$ is the wavenumber in the background medium, and axis $x$ is directed along the tangential component of the wavevector. This  reflectarray method \cite{Berry_reflectarray,Stark_PA,Chang_RA,RA,RA2}  is also known as the  \textit{generalized reflection law} in the optics community \cite{yu2011light}, and it allows controlling the direction of the reflected wave by engineering the reflection phase produced by each small area of the metasurface. It is important to notice that, according to this definition, the local reflection coefficient defines the relation between the incident and reflected fields for each point as if the metasurface were  locally homogeneous, i.e, waves incident at any point are reflected specularly, although with different phases at different points (see Fig.~\ref{fig:Reflection}). For this reason, the use of local reflection coefficient for modelling metasurfaces in far-field scattering calculations is not always accurate, see e.g. \cite{Diaz_From_2017,budhu2020perfectly}.

\begin{figure}[ht]
\centerline{\includegraphics[width=0.9\columnwidth]{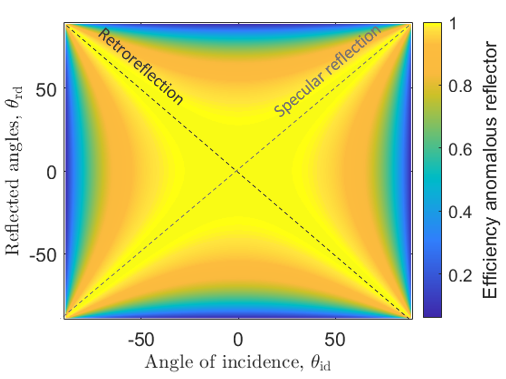}}
\caption{Efficiency of anomalous reflectors characterized by a local reflection coefficient with a linear phase gradient. The efficiency is calculated as $\eta=4\cos{\theta_{\rm id}}\cos{\theta_{\rm rd}}/(\cos{\theta_{\rm id}}+\cos{\theta_{\rm rd}})^2$. Derivation of this formula and a detailed discussion can be found in \cite{asadchy2017eliminating}. }
\label{fig:Efficiency_conventional}
\end{figure}

For better understanding of this issue, we need to analyze the anomalous reflection efficiency, defined as the percentage of energy sent into the desired direction, for metasurfaces designed using a linear variation of the local reflection phase.
{ Parasitic specular reflections from reflectarrays with linear phase gradient  have been noticed in simulations and experiments, e.g., \cite{Rahmat0,Rahmat1}, but the diffraction nature of this effect is not well understood in the antenna community. On the other hand, the diffraction-grating nature of phase-gradient layers is well known in the optical community, e.g., \cite{Magn}, but the known models are limited to the local reflection phase approximation and infinite arrays. 
Actually, phase-gradient} metasurfaces perform high-efficiency anomalous reflection when the difference between the incident and reflected angles is small or when the energy is sent back into the same direction as the incident wave (the retroreflection scenario, $\theta_{\rm rd}=- \theta_{\rm id}$). The energy not sent into the desired direction is coupled to other propagating modes and scattered into other directions, and this phenomenon cannot be modeled if we just use the local reflection coefficient in the far-field calculations. The efficiency of phase-gradient metasurfaces 
is illustrated in Fig.~\ref{fig:Efficiency_conventional}, where one can see that a local reflection coefficient with a linear phase gradient ensures perfect performance around the specular and retroreflection directions. However,
as it was explained and demonstrated in \cite{asadchy2016perfect,Diaz_From_2017,Estakhri_Wavefront_2016}, the efficiency of phase-gradient anomalous reflectors with large transformations of the wave propagation direction drops considerably due to the impedance mismatch between the incident and reflected waves \cite{asadchy2017eliminating}.  Alternative, theoretically perfect  solutions have been found using nonlocal structures \cite{Diaz_From_2017,Epstein_Unveiling_2017}, diffraction gratings \cite{Radi_Metagrating_2018}, auxiliary evanescent fields \cite{epstein2016synthesis}, or power-flow-conformal surfaces \cite{Diaz_Power_2019}. These solutions offer possibilities  for creation of extreme anomalous reflectors with high efficiency and efficient finite-size reflectarrays \cite{budhu2020perfectly}. However, the reflective properties of these advanced metasurfaces cannot be modelled by a local reflection coefficient varying over the reflecting plane.

Current knowledge about metasurface-based anomalous reflectors allows us to understand such important parameters as the efficiency and the energy distribution of the parasitic reflections for the design conditions. However, in order to characterize the propagation channel in environments with integrated metasurfaces, it is necessary to study the response of anomalous reflectors for angles of incidence that are different from the design incidence angle ($\theta_{\rm i}\neq \theta_{\rm id}$). 
In other words, it is necessary to characterize the angular response of metasurfaces for arbitrary illuminations and provide effective  design tools for wireless communication systems and other applications. This is necessary because in any realistic dynamic or multi-path scenario metasurfaces are illuminated from many directions.
In this paper, we study this problem and make a systematic analysis of the angular response of phase-gradient anomalous reflectors designed using the phased-array principle (the generalized reflection law).  

Another important problem that must be solved for further developments of anomalous reflector technology is the estimation of  fields reflected from finite-sized metasurface panels. Current literature on metasurface reflectors focuses on the design and properties of infinite planar metasurfaces illuminated by plane waves, and this knowledge is not enough for finding fields that are reflected and scattered by panels of a  limited size. Known approaches to calculation of reflection from finite-sized anomalous reflectors are based on the local reflection coefficient model, that is, on the assumption that at each point of the metasurface the reflected field equals the incident field  with the desired phase shift, e.g.,  \cite{pozar1997design,basar2019wireless,smart4,smart5,smart7,DiRenzo_arxiv}.  In some works, the amplitude of the reflected field is scaled to ensure that all the incident power is reflected to the desired direction \cite{smart3,smart6}, corresponding to perfect operation. The control of the reflection phase is assumed to be achieved by engineering the local surface impedance of the reflector, using the locally periodic approximation. Similar assumptions are made in modeling finite-size metasurfaces for control of transmission \cite{Francisco,Stefano}.

As we already discussed for the case of infinite metasurfaces, these models do not account for parasitic reflections due to impedance mismatch and the surface periodicity  \cite{Smith,asadchy2016perfect,Diaz_From_2017,Estakhri_Wavefront_2016}.  Another limitation is that the known models can be used only for a single plane-wave illumination at the design incidence angle, and they are not useful if the metasurface is illuminated from many directions (the multi-path scenario). Moreover, the local reflection coefficient model is tantamount to the assumption of locally specular reflection, which is approximately valid only for slowly varying reflecting properties of the metasurface (that is, only when the tilt angles are small). 

Here, we propose an approximate analytical method for calculations of reflected fields in the far zone, merging approaches based on the physical optics and on the theory of diffraction gratings. This model takes into account parasitic, multi-beam scattering and is applicable for arbitrary illumination angles. We hope that by using this method it will become possible to extend the applicability of ray-tracing algorithms for estimation of propagation channel to engineered environments with reconfigurable intelligent metasurfaces (RIS). From the more general perspective, the developed theory allows effective calculation of far-zone scattered fields for general periodical structures that operate as diffraction gratings. 

\section{Problem statement and the definition of mode conversion efficiency}

The first goal of this study is to  analyze the response of anomalous reflectors of infinite extent when they are illuminated from an arbitrary direction, so that the illumination angle, $\theta_{\rm i}$, can be different from the angle of incidence for that the metasurface has been designed or tuned, $\theta_{\rm id}$. Our intention is to explore the response of these devices in the  general way, without considering specific phenomena associated with a particular implementation. For this reason, we model the metasurfaces as boundaries whose surface impedance varies along the reflection plane \cite{Diaz_From_2017,Diaz_Power_2019}.
The surface impedance is defined locally as the ratio between the tangential components of the surface-averaged electric and magnetic fields. The averaging is made on the scale of the unit cells of the metasurface.

Here, for simplicity, we consider planar metasurfaces whose properties vary only along one direction. Axis $x$ of a Cartesian coordinate system is in the metasurface plane and points along the direction of impedance variation. Axis $y$ is normal to the metasurface plane. Such metasurfaces can tilt the reflected plane wave direction in the $xy$-plane. The relation between the local reflection coefficient $\Gamma$, dictated by the phased-array principle (the generalized reflection law), and the surface impedance $Z_{\rm s}$ can be written as $\Gamma(x)=[Z_{\rm s}(x)-Z_{\rm w0}]/[Z_{\rm s}(x)+Z_{\rm w0}]$, where $Z_{\rm w0}$ is the characteristic impedance of the incident plane wave -- the ratio between the tangential to the reflector plane components of the electric and magnetic fields of the incident wave. The impedance of the incident plane wave can be written as $Z_{\rm w0}
^{\rm (TE)}=Z_0/\cos{\theta_{\rm i}}$ for TE-polarized waves and $Z_{\rm w0}
^{\rm (TM)}=Z_0\cos{\theta_{\rm i}}$ for TM-polarized waves. Here,  $Z_0$ is the wave impedance of the surrounding medium (usually air). From this expression, it is straightforward to find the surface impedance of phase-gradient anomalous reflectors as
\begin{equation}
    Z_{\rm s}(x)=jZ_{\rm w0}\cot{[(\sin\theta_{\rm id}-\sin\theta_{\rm rd})kx/2]}\label{eq:surface_impedance}
\end{equation}

The surface impedance is a periodic function of the coordinate $x$, $Z_{\rm s}(x)=Z_{\rm s}(x+D)$, whose period can be found as $D=\lambda/|\sin{\theta_{\rm id}}-\sin{\theta_{\rm rd}}|$, e.g., \cite{Diaz_From_2017}. This period defines the scattering properties of the metasurface and has  important influence on the angular response of periodic metasurfaces. As it is known from the Floquet theory, periodical variations of  metasurface properties lead to existence of multiple diffracted modes which depend on the illumination angle,  the period, and the frequency. Due to the periodicity of the surface impedance, the tangential wave number of different diffracted modes must satisfy
\begin{equation}
    k_{xn}=k\sin{\theta_{\rm i}}+\frac{2\pi n}{D}\label{eq:tangential_WN}
\end{equation}
where $n$ is the order of the diffracted mode. With this notation, mode $n=0$ corresponds to the specular reflection. Consequently, the normal component of the wave vector is $k_{yn}=\sqrt{k^2-k_{xn}^2}$. It is important to note that only when the value of  $n$ ensures that $k>k_{xn}$, the diffracted mode of index $n$ propagates into the far zone.
For propagating modes, the angle of reflection reads
\begin{equation}
    \theta_{{\rm r} n}=\arctan{({k_{xn}}/{k_{yn}})}\label{eq:reflection_angle}
\end{equation}
The incident power is distributed between  the propagating modes depending on the surface impedance of the metasurface. As we can see from Eq.~\eqref{eq:reflection_angle}, the propagation direction of  diffracted modes  strongly depends on the illumination angle, and for different illumination angles the propagation directions of reflected waves  change. Moreover, this dependence is different for different diffracted modes. 

The use of the surface impedance model allows us to numerically study the power distribution among all diffracted modes using the mode-matching method. Recently, this method has been used for the analysis of metasurfaces, allowing fast calculations of  scattering parameters. For a planar metasurface laying in the  $xz$-plane, we can write the tangential components of the electric and magnetic fields in form of the Floquet series:
\begin{eqnarray}
E_t(x)=e^{-jk\sin{\theta_{\rm i}}x}+\sum_{n=-\infty}^{\infty}A_n e^{-jk_{xn}x}\label{Ef}
\\
H_t(x)=\mp \left[Y_{{\rm w}0}e^{-jk\sin{\theta_{\rm i}}x}-\sum_{n=-\infty}^{\infty}A_n Y_{{\rm w}n} e^{-jk_{xn}x}\right] \label{eq:magnetic_field}
\end{eqnarray}
where $A_n$ represents the amplitude coefficient for the $n$-th  reflected harmonic. Notice that the amplitude coefficients are normalized to the amplitude of the tangential component of the incident field. 
The `$\mp$' sign in Eq.~\eqref{eq:magnetic_field} is  negative for TE-polarized waves and positive for TM-polarized waves. 
The wave admittance for the diffracted modes can be calculated as $Y_{{\rm w}n}
^{\rm (TE)}=k_{zn}/\omega\mu_0$ for TE-polarized waves and  $Y_{{\rm w}n}
^{\rm (TM)}=\omega\varepsilon_0/k_{zn}$ for TM-polarized waves.
By enforcing the impedance boundary condition  $\mathbf{E}_t(x)=Z_{\rm s}(x)[\hat{\mathbf{y}}\times \mathbf{H}_t(x)]$, where $\hat{\mathbf{y}} $ is the unit normal vector, we determine the amplitudes of all Floquet harmonics. 
Full description of this method can be found in  \cite{hwang2012periodic} and a MATLAB code that implements this mode-matching method for metasurfaces modeled by space-varying surface impedance can be found in  \cite{wang2018extreme}. 

Knowing the amplitudes of all the reflected waves, one can easily calculate the conversion efficiency for each propagating plane-wave mode,  defined as the ratio between the normal component of the reflected power curried by a certain mode and the normal component of the  incident power: $\eta_n=P_{yn}/P_{y}^{\rm in}$, where $P_{yn}=(1/2)\, \Re{(\mathbf{E}_{tn}\times \mathbf{H}_{tn}^*)}$, and $P_{y}^{\rm in}$ is the amplitude of the normal component of the Poynting vector of the incident plane wave. For TE-polarized incidence, the power carried away from the metasurface (per unit area) by each Floquet  harmonic can be expressed in terms of the corresponding mode-conversion efficiencies  
\begin{equation}
    \eta_n^{(\rm TE)}=
    \begin{cases}
      {|A_n|^2}\frac{\cos\theta_{{\rm r}n}}{\cos{\theta_{\rm i}}} &  \text{for propagating modes} \\
      0      &  \text{for evanescent modes}
    \end{cases}
    \label{eff_TE}
\end{equation}
The conversion efficiency for each diffracted mode for TM-polarized waves is defined as 
\begin{equation}
    \eta_n^{(\rm TM)}=\begin{cases}
     {|A_n|^2}\frac{\cos{\theta_{\rm i}}}{\cos\theta_{{\rm r}n}} &  \text{for propagating modes} \\
      0      &  \text{for evanescent modes}
    \end{cases}
    \label{eff_TM}
\end{equation}
For lossless metasurfaces the sum of the conversion efficiencies of all propagating modes equals unity.

\section{Angular Properties of Anomalous Reflectors}
\label{sec:angular}

In this section, we will study how the energy is scattered when the anomalous reflectors are illuminated at illumination angles different from the design angle of incidence, i.e., we study  the angular response of the metasurface.  Before continuing with this analysis, it is necessary to define the surface impedance for different illumination angles. In this study, for simplicity, we will assume that the surface impedance defined in Eq.~\eqref{eq:surface_impedance} remains constant for any illumination. This means that we neglect spatial dispersion of the metasurface response \cite{luukkonen2009effects,Mario_2008,modeboo} or, in other words, the dependence of the surface impedance on the tangential wavevector.  In actual metasurfaces, this assumption is approximately valid if we choose a proper structure for implementing the metasurface. For example, a metasurfaces based on metallic grooves acting as local phase shifters or metallic patterns over a high-permittivity dielectric layer backed by a metallic plane fulfill this requirement.  Section~VI provides the analysis of an actual structure which satisfy Eq.~\eqref{eq:surface_impedance} for any angle of incidence.

%Before continuing with the analysis of the angular response, it is necessary to define the surface impedance for different illumination angles. In this study, for simplicity, we will assume that the surface impedance remains constant for any illumination angle. This means that we neglect spatial dispersion of the metasurface response.  This assumption is approximately valid if we choose a proper topology for the actual implementation of the metasurface. For example, one can consider metasurfaces based on metallic grooves acting as local phase shifters or metallic patterns over a high-permittivity dielectric layer backed by a metallic plane. More details about the effect of the topology on the angular response will be presented in Section~\ref{sec:topology}.

\begin{figure}[ht]
\subfigure[]{\includegraphics[width=.9\linewidth]{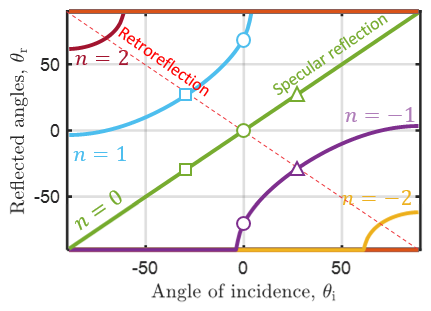}}
\subfigure[]{\includegraphics[width=.9\linewidth]{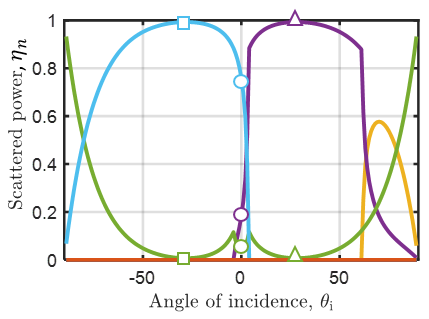}}
\caption{Analysis of the scattering properties of a flat  infinite anomalous reflector designed for $\theta_{\rm id}=0^\circ$ and $\theta_{\rm rd}=70^\circ$. (a) For a periodicity given by $D=\lambda/|\sin{\theta_{\rm id}}-\sin{\theta_{\rm rd}}|$, relation between the angle of incidence $\theta_{\rm i}$ and the angle of reflection of the diffracted modes $\theta_{\rm r}$. (b) Energy distribution between the reflected propagating harmonics for different incident angles. The power is normalized to the amplitude of the normal component of the incident power flow. }
\label{fig:Conventiona_0_70}
\end{figure}

We start the study of the angular response with a phase-gradient anomalous reflector designed for $\theta_{\rm id}=0^\circ$ and $\theta_{\rm rd}=70^\circ$. This is a typical example of a metasurface designed for a large tilt of the reflected wave direction (with respect to the usual specular-reflection angle).  The period of this reflector is $D=1.0642\lambda$. The first step in the analysis is to determine what diffracted modes propagate in the system for different illumination angles. 
Figure~\ref{fig:Conventiona_0_70}(a) shows the directions of propagation of the diffracted modes for different illumination angles. From these results we can see that only diffracted modes with the orders $n\in[-2,2]$ contribute to the angular response in the far zone (even though many other modes can be exited, they do not propagate away from the surface and contribute only to the reactive fields in the vicinity of the reflecting surface). 

Next, we study how the energy is distributed between all these propagating modes for different illumination angles. The energy distribution for different illumination angles is represented in Fig.~\ref{fig:Conventiona_0_70}(b). For the design conditions, i.e., for the normal illumination,  most of the incident power (76\%) goes to the first diffracted mode $n=1$ [see circular symbols in Fig. \ref{fig:Conventiona_0_70}], realizing the desired anomalous reflection. For this illumination angle, as it is expected from the design conditions, this mode propagates at $\theta_{\rm r}=70
^\circ$. The rest of the incident power  is scattered toward  $\theta_{\rm r}=-70
^\circ$ direction (mode $n=-1$) and in the  normal direction (mode $n=0$). These results agree with the previous analysis of anomalous reflectors  done in \cite{asadchy2017eliminating,Diaz_From_2017}. The energy distribution in Fig.~\ref{fig:Conventiona_0_70}(b) shows that when the angle of incidence $\theta_{\rm i}\in [-70, 70] 
^\circ$, most of the incident energy is transferred to higher-order modes ($n\neq 0$). For angles of incidence  $|\theta_{\rm i}|> 70 
^\circ$, specular reflection becomes dominant.

It is also important and instructive to consider the case when this metasurface is illuminated from $\theta_{\rm i}=-28
^\circ$ (see square symbols in Fig.~\ref{fig:Conventiona_0_70}). For this specific illumination angle, all the energy is efficiently transferred to the diffracted mode $n=1$. To understand this behavior, one needs to notice that at this angle of incidence, the propagation direction of this diffracted mode  corresponds to the retroreflection scenario. As we can see from Fig.~\ref{fig:Efficiency_conventional}, retroreflection is the only nontrivial transformation of the reflected waves that produces perfect performance when we use a linear phase gradient of the reflection phase. Equation~\eqref{eq:surface_impedance} can be written for the retroreflection case as
$Z_{\rm s}(x)=jZ_w\cot{(\sin\theta_{\rm retro}kx)}$, with $\theta_{\rm retro}$ being the angle of retroreflection. As it has been demonstrated in \cite{Asadchy_Flat_2017}, the boundary condition for this surface impedance is perfectly satisfied by the sum of only two plane waves: the incident and the retroreflected ones, ensuring  perfect performance.
 By a simple comparison of this equation and Eq.~\eqref{eq:surface_impedance}, we can see that the retroreflection angle and the reflection angle in the design of the anomalous reflector are related as
\begin{equation}
  \theta_{\rm retro}=\arcsin{(\sin\theta_{\rm id}-\sin\theta_{\rm rd})/2}\label{eq:angle_retro}
\end{equation}
For the specific case when $\theta_{\rm id}=0^\circ$ and $\theta_{\rm rd}=70^\circ$, the retroreflection angle calculated using this expression equals  $\theta_{\rm retro}=-28^\circ$. As under our assumptions the surface impedance remains the same  for different illumination angles, an anomalous reflector designed for $\theta_{\rm id}=0^\circ$  and $\theta_{\rm rd}=70^\circ$ performs perfect retroreflection when the surface is illuminated at $\theta_{\rm i}=-28^\circ$.

Another interesting aspect of the angular response appears when the illumination comes from  $\theta_{\rm i}=28^\circ$ (see triangular symbols in Fig.~\ref{fig:Conventiona_0_70}).  Here, we can see that all the incident power is transferred to the diffracted mode $n=-1$, meaning that all the power is sent back into the same direction. In other words, an anomalous reflector designed for $\theta_{\rm id}=0^\circ$  and $\theta_{\rm rd}=70^\circ$ also performs perfect retroreflection when $\theta_{\rm i}=28^\circ$.  The explanation of this behaviour is simple. As the metasurface is reciprocal, specular reflection phenomenon is  symmetric with respect to the illumination angle sign [see the green line in Fig.~\ref{fig:Conventiona_0_70}(b)]. This means that,  as no power is reflected specularly when the metasurface is illuminated from $\theta_{\rm i}=-28^\circ$, no power will be sent into the specular direction when $\theta_{\rm i}=28^\circ$. The period of this metasurface determines that only one diffracted mode propagates when $\theta_{\rm i}=\pm 28^\circ$ and, considering that specular reflection is not allowed by reciprocity, the power  must go to the diffracted mode $n=-1$ to satisfy the conservation of power.

\begin{figure}[ht]
\subfigure[]{\includegraphics[width=.9\linewidth]{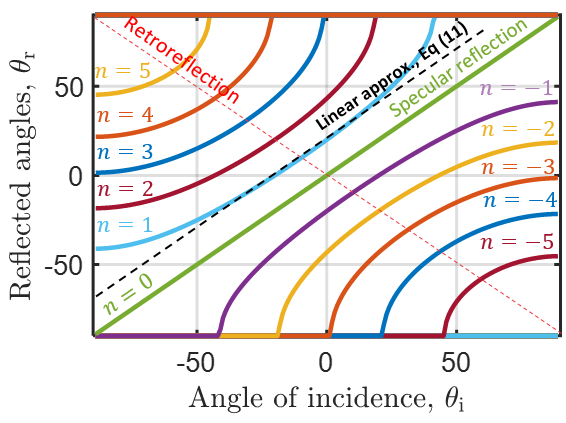}}
\subfigure[]{\includegraphics[width=.9\linewidth]{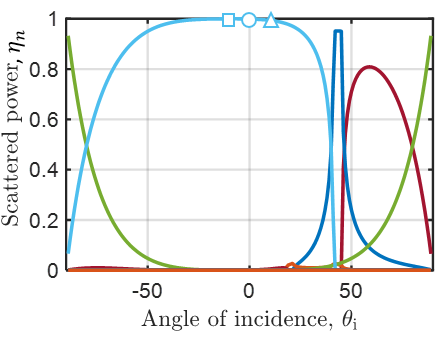}}
\caption{Analysis of the scattering properties of an infinite anomalous reflector designed for $\theta_{\rm id}=0^\circ$ and $\theta_{\rm rd}=20^\circ$. (a) For a periodicity given by $D=\lambda/|\sin{\theta_{\rm id}}-\sin{\theta_{\rm rd}}|$, relation between the angle of incidence $\theta_{\rm i}$ and the angle of reflection of the diffracted modes $\theta_{\rm r}$. The dashed line corresponds to the linear approximation \eqref{linear}. (b) Power distribution between the propagating harmonics for different incident angles.}
\label{fig:Conventiona_0_20_infinite}
\end{figure}

The second case of study is an anomalous reflector that is designed for $\theta_{\rm id}=0^\circ$ and $\theta_{\rm rd}=20^\circ$. This is a typical example of a small tilt of reflection direction, when the impedance is a slowly varying periodical function of $x$, and the period is electrically large. In this specific example, the period of the metasurface is $D=2.9\lambda$. As in the previous case, to understand the scattering properties of this metasurface, it is necessary to consider what  diffracted modes propagate in the system for different incident angles. Figure~\ref{fig:Conventiona_0_20_infinite}(a) shows the reflection angles for different illuminations and for each propagating mode. The most evident difference with the previous example is that, due to the larger period, there are more propagating modes. We can also see that the dependence  of the reflection angle on the incident angle becomes more linear. It is particularly interesting to see that for the first diffracted mode, $n=1$, the dispersion can be approximated by the linear relationship 
\begin{equation}
    \theta_{\rm r}=\theta_{\rm i}+ 2|\theta_{\rm retro}|
    \label{linear}
\end{equation}
where $\theta_{\rm retro}=-9.5
^\circ$. This linear approximation is 
shown in 
Fig.~\ref{fig:Conventiona_0_20_infinite}(a) as a dashed line.

The reflected power distribution is represented in Fig.~\ref{fig:Conventiona_0_20_infinite}(b) for different illumination angles showing important differences in comparison to the previous example. Here, we can see that the efficiency of the anomalous reflector for the design conditions (see circular symbols) increases, and nearly all the power  illuminating the metasurface from normal direction goes into the desired direction, $\theta_{\rm r}=20^\circ$.
However, when the metasurface is illuminated from the retroreflection angle, $\theta_{\rm i}=-9.5
^\circ$, all the power is sent back (see square symbols). 

%The radiation pattern of a finite metasurface design with these conditions can be estimated by Eq.\ref{eq:radiation_pattern}. Figure \ref{fig:Conventiona_0_20_finite} represents the radiations pattern under three different illuminations.

\begin{figure}[ht]
\includegraphics[width=1\linewidth]{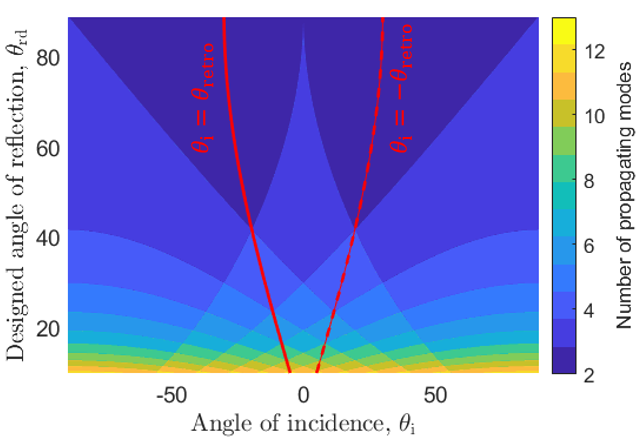}
\caption{The number of propagating diffracted modes for anomalous reflectors designed for $\theta_{\rm id}$. Red line marks the incident angles for perfect retroreflection $\theta_{\rm i}=\theta_{\rm retro}$. Dashed red line marks the incident angles for the reciprocal scenario of perfect retroreflection $\theta_{\rm i}=-\theta_{\rm retro}$. }
\label{fig:NumberModes}
\end{figure}

An important difference appears when we illuminate the metasurface from the opposite direction,  $\theta_{\rm i}=9.5
^\circ$. In this case, all the power is still transferred to the diffracted mode $n=1$ that sends the power into $\theta_{\rm r}=31.5
^\circ$ (see triangular symbols). With this configuration, the existence of multiple diffracted modes allows satisfying the reciprocity condition of the system and the conservation of power without enforcing perfect retroreflection for this illumination. From this result, we can conclude that only in the cases where two modes propagate in the system  (specular reflection and an additional diffracted mode), retroreflection is enforced by reciprocity when $\theta_{\rm i}=-\theta_{\rm retro}$. Figure~\ref{fig:NumberModes} presents the results of the analysis of the number of propagating modes for anomalous reflectors designed for the normal illumination. Here we can see that perfect retroreflection due to reciprocity only takes place when $\theta_{\rm rd}>42
^\circ$.

\begin{figure}[ht]
\includegraphics[width=1\linewidth]{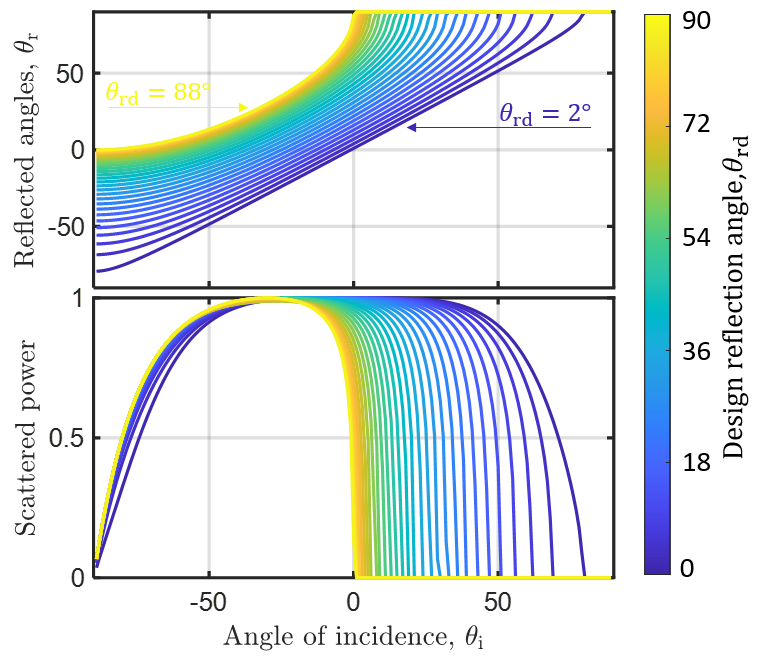}
\caption{Angular response of the first diffracted mode, $n=1$, for different $\theta_{\rm rd}$ when $\theta_{\rm id}=0$. (a) Angular dispersion for different design conditions. (b) Normalized scattered power by the diffracted mode.}
\label{fig:Comparison_n1}
\end{figure}

It has been demonstrated that the first diffracted mode, $n=1$, significantly contributes to scattering  the impinging power. The upper panel in Fig.~\ref{fig:Comparison_n1} presents the angular dispersion of the diffracted $n=1$ mode for different illumination angles and different design conditions. The angular dispersion of this mode that defines the direction where the energy is scattered, $\theta_{\rm r}$, and the amount of power carried by this mode depend on 
the design conditions (the angles $\theta_{\rm id}$ and $\theta_{\rm rd}$). Here, we can see that keeping $\theta_{\rm id}=0$, the angular dispersion curve of the mode becomes wider and more linear when $\theta_{\rm rd}$ decreases. Importantly,  the impinging power predominately scatters into the first mode when the angular dispersion is approximately linear, as we can see in the bottom panel of Fig.~\ref{fig:Comparison_n1}.  This is because the linear phase gradient of the reflection coefficient introduces the proper additional linear momentum to produce this transformation.

We see that phase-gradient anomalous reflectors function as diffraction gratings, scattering incident power into different propagating modes of the Floquet expansion depending on the illumination angle. The linear phase gradient optimized for a specific pair of the incidence and reflection angles ensures maximization of the power reflected into the desired direction, but ideal conversion efficiency for phase-gradient metasurfaces is possible only for retrodirection and for trivial specular reflection. When the incidence angle changes, reflective properties change and the functionality of the metasurface changes (from high-efficiency anomalous reflection to retroreflection).

\section{Effective Reflection Coefficients and Far-Field Scattering}
\label{sec:rn}

As we have seen in the previous sections, for the analysis and design of phase-gradient metasurfaces, it is convenient to use the local reflection coefficient, defined as the ratio between the tangential component of the incident and reflected electric fields at each point of the metasurface. This reflection coefficient models \textit{local} properties of the reflector (averaging is made on the scale of one meta-atom) and allows the calculation of the surface impedance. However, as it was explained in the introduction, this parameter does not allow proper  modeling of reflection from the metasurface as a whole and understanding of the effects on the  propagation in environments with metasurfaces. The main reason is that gradient metasurfaces are diffraction gratings which scatter into several directions. In the calculation of far-field scattering, it is important to define the reflection coefficient of infinite metasurfaces as the ratio between the reflected electric fields of the  plane-wave Floquet harmonics and the incident electric field, or, in other words, characterize the \textit{macroscopic} behavior of the metasurface  according to Eq.~(\ref{eq:macroscopic}).
%Typically, if we consider plane-wave reflection, the reflection coefficient is defined as the ratio between the tangential components of the electric field of the incident plane wave and the reflected plane wave. In the case of metasurfaces, the existence of multiple  reflected plane waves does not allow us to use this definition and introduce a  unique reflection coefficient. 
%For this reason,  a macroscopic reflection coefficient for modeling metasurfaces can  be defined as a combination of  the individual reflection coefficients for each individual reflected mode as   
%\begin{equation}
%    R(\theta_{\rm i},x)=\sum_nr_n(\theta_{\rm i})e^{-jk(\sin\theta_{{\rm r}n}-\sin\theta_{\rm i})x}e^{-jk(\cos\theta_{{\rm r}n}+\cos\theta_{\rm i})y}  
%\end{equation}
%where $r_n(\theta_{\rm i})=|r_n(\theta_{\rm i})|\exp{[j\phi_n(\theta_{\rm i})]}$ is the individual reflection coefficient for each propagating diffracted mode. The angles $\theta_{{\rm r}n}$ defined by
%\begin{equation}
%\sin\theta_{{\rm r}n}=\sin\theta_{\rm i}+\frac{n2\pi}{kD}
%,\quad \cos\theta_{{\rm r}n}=\frac{\sqrt{k^2-(\sin\theta_{\rm i}+n2\pi/D)^2}}{k^2}
%\label{trn}
%\end{equation}
%are the reflection angles of the propagating harmonics of index $n$. 

Using Eq.~\eqref{Ef}, we can define the value of the reflection coefficient in terms of the normalized amplitude of the reflected field, as  $r_n(\theta_{\rm i})=A_n$. 
%Here, the reflection coefficient is defined for an arbitrary distance from the metasurface. At the reflecting surface, we let $y=0$.
Here, we consider only the propagating Floquet harmonics. Only these modes  contribute to the macroscopic, far-field  response of  metasurfaces of infinite extent (the far-field reflection coefficient for evanescent fields is zero).
Evanescent harmonics of the fields contribute to reflections from finite-sized metasurfaces due to edge scattering effects, and we will discuss this issue later. We note that macroscopic reflection coefficients in the assumption that the metasurface ideally reflects only one plane wave were discussed earlier in \cite{smart6}.

\begin{figure}[ht]
\includegraphics[width=1\linewidth]{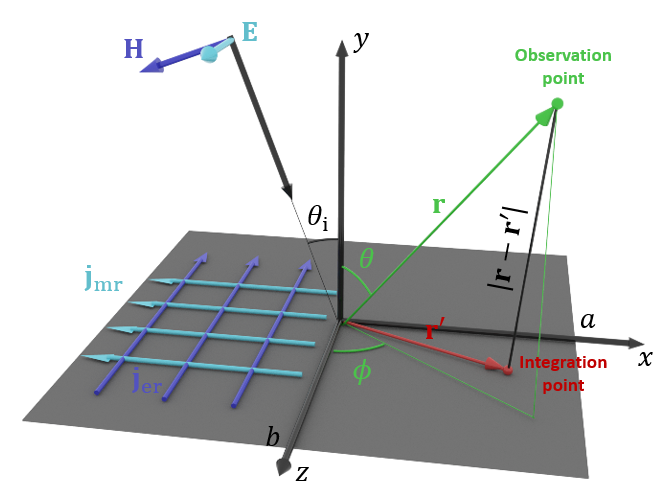}
\caption{Geometry of the problem. A metasurface of rectangular shape and a finite size  ($2a\times 2b$) is placed on  the $xz$-plane. The illumination is considered to be a TE-polarized plane wave, and the incidence plane is the $xy$-plane.}
\label{fig:geometry}
\end{figure}

This analysis is a useful tool when we consider infinite metasurfaces that extend over a whole plane. However, for realistic scenarios with finite-size metasurfaces we need to consider the scattering pattern of the metasurface. If we assume that the metasurfaces under study are large compared to the free-space wavelength and that the unit-cell size is small compared to the wavelengths of field variations along the metasurface plane $\lambda_n=2\pi/k_{xn}$ for all significant modes, we can use the physical-optics approximation. Namely, we can assume that the currents induced on the finite-size metasurface are the same as on the infinite metasurface. That is, we  neglect the perturbations of induced currents near the edges of the metasurface. Importantly, note that we do not assume a local relation between the incident field and the induced currents, which is the other conventional assumption of the physical optics, see, e.g., \cite{Osipov}. Instead, we find the induced currents solving the reflection problem for a periodically modulated impedance boundary, which results in the knowledge of the macroscopic reflection coefficients $r_n$. 

For simplicity of writing, we assume that the metasurface has a rectangular shape, occupying  the area defined by $-a<x<a$ and $-b<z<b$ with $2a$ and $2b$ being the total sizes of the metasurface. The geometry of the problem is shown in Fig.~\ref{fig:geometry}. 
If the metasurface is illuminated by a TE-polarized wave defined by the electromagnetic field $\mathbf{E}_{\rm i}=E_0 e^{-jk(\sin\theta_{\rm i}x-\cos\theta_{{\rm i}n}y) }\hat{\mathbf{z}}$, the reflected electric field at a certain point of  the metasurface plane (position vector $\mathbf{r'}$ with the  coordinates $x',z'$)  reads 
\begin{equation}
      {E}_{{\rm r}z}=E_0 \sum_nr_n(\theta_{\rm i})e^{-jk\sin\theta_{{\rm r}n}x'} 
\end{equation}
Here, we use the physical-optics approximation, assuming that the reflected field is not disturbed near the metasurface edges.
The components of the magnetic field associated with the reflected electric field can be written as 
\begin{eqnarray}
      {H}_{{\rm r}x}=E_0Y_0 \sum_nr_n(\theta_{\rm i})\cos\theta_{{\rm r}n}e^{-jk\sin\theta_{{\rm r}n}x'} \\
      {H}_{{\rm r}y}=-E_0Y_0 \sum_nr_n(\theta_{\rm i})\sin\theta_{{\rm r}n}e^{-jk\sin\theta_{{\rm r}n}x'} 
\end{eqnarray}
with $Y_0=k/\omega\mu$ being the admittance of the background medium. Note that this relation is written for plane waves propagating along the directions defined by the reflection angles $\theta_{{\rm r}n}$, because we do not assume locally specular reflection.

Using the Huygens principle, we introduce equivalent electric and magnetic surface currents that create these reflected fields. The corresponding surface current densities read  
$\mathbf{j}_{\rm er}= \hat{\mathbf{y}}\times\mathbf{H}_{\rm r}$ and $\mathbf{j}_{\rm mr}=-\hat{\mathbf{y}}\times\mathbf{E}_{\rm r}$. 
On the shadow side we assume, within the physical optics approximation, that the total field is zero, since the metasurface is impenetrable. Thus, the fields scattered by the metasurface in the shadow direction are equal to the negative of the incident fields. The corresponding equivalent Huygens shadow surface current densities read $\mathbf{j}_{\rm es}= \hat{-\mathbf{y}}\times(-\mathbf{H}_{\rm i})$ and $\mathbf{j}_{\rm ms}=\hat{\mathbf{y}}\times(-\mathbf{E}_{\rm i})$. Since the electrical thickness of the metasurface is negligible, we can sum up these two Huygens currents and assume that both flow at the same surface. The total Huygens current densities correspond to the total fields at the illuminated side:   
\begin{equation}
    \mathbf{j}_{\rm e}= \hat{\mathbf{y}}\times(\mathbf{H}_{\rm i}+\mathbf{H}_{\rm r}),\quad \mathbf{j}_{\rm m}=-\hat{\mathbf{y}}\times(\mathbf{E}_{\rm i}+\mathbf{E}_{\rm r} )
    \label{Hc}
\end{equation}

In a previous work \cite{smart3}, it was assumed that phase-gradient reflectors can be viewed as infinitely thin sheets of only electric surface currents with appropriately modulated phase, but that assumption contradicts to the assumption of an impenetrable reflector. Indeed, if the metasurface carries only an electric surface current, the tangential electric field is continuous across the metasurface. Thus, if the field behind the metasurface is zero, the only allowed value of the reflection coefficient is $-1$, and no phase modulation is possible. For this reason, consistent models  require introduction of both electric and magnetic surface currents. Magnetic surface current density can be expressed in terms of the electric current using the characteristic impedance \emph{of the corresponding reflected mode}. Note that the conventional physical optics expressions (for example, for calculations of far fields of aperture antennas) assume the free-space impedance relation between the electric and magnetic fields at the radiating surface, which in our case corresponds to the assumption of locally specular reflection.  This assumption is valid only for slowly varying surface impedances, and its use cannot be justified for majority of anomalously reflecting metasurfaces.

For the equivalent Huygens surface currents in form \eqref{Hc}, the reflected electric field at any point in space (position vector $\mathbf{r}$) is given by the following integral (e.g., \cite[Eq.~(3.87)]{Osipov})
\begin{equation}
     \mathbf{E}_{\rm sc}(\mathbf{r})=
     {1\over j\omega \varepsilon}
[\nabla\nabla+k^2]\int_S   \mathbf{j}_{\rm e}(\mathbf{r'})G_0( \mathbf{r},\mathbf{r'})dS'
\nonumber\end{equation}
\begin{equation}
-\nabla\times
\int_S   \mathbf{j}_{\rm m}(\mathbf{r'})G_0( \mathbf{r},\mathbf{r'})dx'dz'
\end{equation}
where 
\begin{equation}
    G_0( \mathbf{r},\mathbf{r'})={1\over 4\pi}{\frac{e^{-jk|\mathbf{r}-\mathbf{r}'|}}{|\mathbf{r}-\mathbf{r}'|}}
\end{equation}
is the scalar Green function, and the integration is over the metasurface plane. This formula can be used for calculations of scattered field at any distance from the metasurface, as long as it is large compared to the unit-cell (meta-atom) size. As is well known from the theory of diffraction, calculations of fields in the far zone can be significantly simplified. In this work, we use the simplified far-field model that is valid under the following assumptions: 
$|\mathbf{r}|\gg \lambda$, $|\mathbf{r}|\gg L$, and $L^2/|\mathbf{r}| \ll \lambda $. Here, $L={\rm max}(2a,2b)$ is the largest size of the metasurface plate. That is, the distance to the observation point is large compared to the wavelength and to the size of the metasurface plate.
Under these assumptions, the reflected field can be found as
\begin{equation}
    \mathbf{E}_{\rm sc}(\mathbf{r})\approx {jk\over 4\pi }
    {\frac{e^{-jk|\mathbf{r}|}}{|\mathbf{r}|}}
    \int_S e^{jk\hat{\mathbf{r}}\cdot \mathbf{r'}}\hat{\mathbf{r}}\times [Z_0\hat{\mathbf{r}}\times \mathbf{j}_{\rm e}(\mathbf{r'})+\mathbf{j}_{\rm m}(\mathbf{r'})]dx'dz'
\end{equation}
where $Z_0=1/Y_0$ is the impedance of the surrounding space and $\hat{\mathbf{r}}$ is the unit vector along $\mathbf{r}$, pointing from the center of the metasurface plate to the observation point. Details of the derivation of this formula and justifications of the simplifying assumptions can be found in \cite[Sec.~3.4.1]{Osipov}.

We see that the far-zone field is a spherical wave, whose fields are orthogonal to vector $\hat{\mathbf{r}}$ (to the direction of scattering). The amplitude and angular dependence of the scattered field are given by the surface integral that can be viewed as the spatial Fourier transform of the surface current distribution. Calculation of this integral is trivial because the integrand is a sum of simple exponential functions. The result is obviously a sum of products of two sinc functions (${\rm sinc}(x)=\sin(x)/x$).

It is convenient to express the final result using spherical coordinates, writing $\hat{\mathbf{r}}=\sin\theta\cos\phi\hat{\mathbf{z}}+\sin\theta\sin\phi\hat{\mathbf{x}}+\cos\theta\hat{\mathbf{y}}$.
Let us consider the scattered fields in the incidence plane ($\phi=\pi/2$). In this case, the $\theta$-component of the reflected field is zero, one of the two sinc functions equals unity,  and the expression for the $z$-component of the scattered electric field can be simplified as 
%\begin{eqnarray}
 %   {E}_{{\rm r} \phi}=jk[A_{{\rm e}z}+Z_0A_{{\rm m} x}\cos\theta]
%\end{eqnarray}
%Using Eqs. (\ref{eq:vector_potential_m_cartesian}) and (\ref{eq:vector_potential_e_cartesian}), the final expression for the reflected field in the plane of incidence can be written as
$$
    {E}_{{\rm sc} z}={jk\over 4\pi }
    {\frac{e^{-jk|\mathbf{r}|}}{|\mathbf{r}|}}E_0S  \left[(\cos\theta-\cos\theta_{\rm i}){\rm sinc}(ka_{\rm ef})\right.$$
    \begin{equation}
    \left.+\sum_nr_n(\theta_{\rm i})(\cos\theta+\cos\theta_{{\rm r}n}){\rm sinc}(ka_{{\rm ef}n})\right]
    \label{great}
\end{equation}
where $a_{{\rm ef}}=(\sin\theta-\sin\theta_{{\rm i}})a $, and   $a_{{\rm ef}n}=(\sin\theta-\sin\theta_{{\rm r}n})a $
represents the effective size of the metasurface for each reflected propagating mode. Furthermore, $S=4ab$ is the area of the metasurface. As a check, for a PEC plate $r_{0}=-1$ for any $\theta_{\rm i}$, and the other terms in the summation equal zero. Furthermore, the reflection angle $\theta_{{\rm r}0}=\theta_{\rm i}$ (specular reflection). In this trivial case, formula \eqref{great} reduces to the well-know result of the physical-optics model of scattering from electrically large PEC plates (e.g., Eq.~(4) in \cite{smart3}   and Eqs.~(9.96) and (9.312) in \cite{Osipov}).

\begin{figure}[ht]
\center
\includegraphics[width=.7\linewidth]{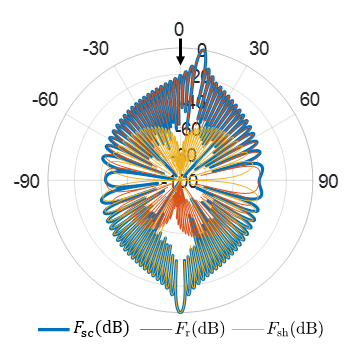}
\caption{Analysis of the scattering properties of an anomalous reflector designed for $\theta_{\rm id}=0^\circ$ and $\theta_{\rm rd}=10^\circ$ under normal illumination. The radiation patterns are represented in dB:  $F_{i}(\rm dB)=20\log_{10}(|F_{i}(\theta)|)$, where $i={\rm sc}, {\rm r}, {\rm sh}$ refer, respectively, to the total scattered field, the field scattered by the equivalent surface currents defined by the reflected fields $\mathbf{E}_{\rm r}$, $\mathbf{H}_{\rm r}$, and the shadow currents defined by the incident fields.}
\label{fig:Radiation_diagram}
\end{figure}

The first term in Eq.~\eqref{great} represents the contribution of the currents generating shadow radiation. In the ideal case of an infinite impenetrable metasurface, the shadow radiation will be zero in the backscattering direction, and it will ensure complete cancellation of the incident field in the forward direction. However, due to the finite size of actual metasurfaces, the contribution of the shadow radiation to the total radiation should be included because these currents create fields in the whole space.
For a clear representation, it is convenient to normalize the pattern so that its maximum value for the PEC plate of the same size is unity. With this normalization, the patterns of the reflected and the shadow fields can be written as
\begin{eqnarray}
    {F}_{{\rm r} }(\theta)=\frac{1}{2\cos\theta_{\rm i}}\sum_nr_n(\theta_{\rm i})(\cos\theta_{{\rm r}n}+\cos\theta){\rm sinc}(ka_{{\rm ef}n})\label{20}\\
    {F}_{\rm sh}(\theta)=\frac{1}{2\cos\theta_{\rm i}}(\cos\theta-\cos\theta_{\rm i}){\rm sinc}(ka_{\rm ef})
\end{eqnarray}
and the total scattering pattern is expressed as $ {F}_{{\rm sc} }(\theta)= {F}_{{\rm r} }(\theta)+ {F}_{{\rm sh} }(\theta)$. 
Figure~\ref{fig:Radiation_diagram} shows the patterns for the reflected, shadow, and total scattered fields produced by an anomalous reflector designed for $\theta_{\rm id}=0^\circ$ and $\theta_{\rm rd}=10^\circ$ under normal illumination. The size of the metasurface is assumed to be $10\lambda \times 10\lambda$. As expected, total backscattering radiation is predominately in $\theta=10^\circ$ and additional secondary lobes appear due to the finite size of the metasurface. From the figure, it is clear that the contribution of the shadow radiation to the total backscattering is only noticeable in the side lobes, where it is significant.

 We note that the reflection pattern of finite-sized anomalous reflectors was recently considered in \cite{DiRenzo_arxiv}. However, the model in that paper does not take into account the diffraction-grating nature of such metasurfaces (only one reflected plane-wave harmonic is assumed, defined by a local reflection coefficient) as well as the shadow-current contribution to scattering; compare with Eq.~(32) in \cite{DiRenzo_arxiv}.

Formula \eqref{great} is general, applicable for both propagating and evanescent harmonics of the reflected fields. Indeed, higher-order, evanescent components of the equivalent currents are also exponential functions of $x'$, and integrals of these components also give sinc functions of the same form as for the contributions from the propagating Floquet harmonics. The difference is that for the evanescent modes the arguments of these sinc functions are large compared to unity. Obviously, for the propagating harmonics the value of $|\sin\theta_{{\rm r}n}| $ is smaller than unity, and the argument of the sinc function can take zero value, corresponding to the main-lobe maximum. However, for evanescent harmonics  of order $n$, the argument of the sinc function 
\begin{equation}
    ka_{{\rm ef}n}=ka\left[(\sin\theta-
   \sin\theta_{\rm i})-n{\lambda\over D}\right]
\end{equation}
does not cross zero for any observation angle $\theta$ because $\left|\sin\theta_{\rm i}+n{\lambda\over D}\right|>1$. {For evanescent modes, the factor $\cos \theta_{{\rm r}n}$ in \eqref{20} becomes imaginary and grows with increasing mode number $n$, but the corresponding mode amplitudes $A_n=r_n$ decrease at approximately the same rate.
Thus, relative contributions of propagating and  evanescent modes of the induced surface currents to scattering in the main-lobe directions are determined by the sinc function amplitude. We see that the contributions of evanescent modes scale as $1/(ka)$
as compared to the contributions of the  propagating harmonics. }
We also note that, for high-order modes, $n{\lambda\over D}$ is a large number as compared to unity. Thus, the corresponding sinc functions weakly depend on the angles of incidence and observation. This means that contributions from evanescent current harmonics affect mainly the side-lobe level and shape, and they are predominantly of diffuse character. Finally we note that, within the physical optics approximation, we can take these modes into account only if the variations of the induced currents over the metasurface plane remain slow on the micro-structure scale, see Section~\ref{sec:topology}. This assumption cannot be justified in the frame of the surface-impedance model. For these reasons, in all numerical examples presented in this paper we include the contributions of propagating field harmonics only, neglecting the small effects of additional diffuse scattering due to edge inhomogeneities  of the evanescent fields. 

\begin{figure}[h]
\includegraphics[width=1\linewidth]{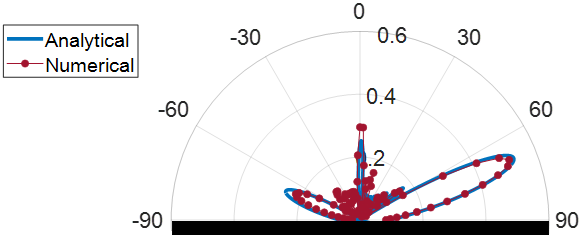}
\caption{Comparison  between the scattered fields calculated using a full-wave commercial software (red symbols) and the approximate method (blue line) when $\theta_{\rm id}=0^\circ$, $\theta_{\rm rd}=70^\circ$, $\theta_{\rm i}=0^\circ$, and $a=b=5D$ .}
\label{fig:Finite_numerical}
\end{figure}

To verify the validity of the proposed approximate model, we conducted numerical simulations of finite-sized metasurfaces modeled with the surface impedance profile dictated by Eq.~\eqref{eq:surface_impedance}. As an example, Fig.~\ref{fig:Finite_numerical} shows the comparison between the scattered fields calculated using a full-wave commercial software \cite{HFSS_software} and using Eq.~\eqref{great} for $\theta_{\rm id}=0^\circ$, $\theta_{\rm rd}=70^\circ$, $\theta_{\rm i}=0^\circ$, and $a=b=5D$. Numerical results confirm the existence of multiple-beam scattering as it is  predicted by our approximate model. The results are in good agreement with the approximate analytical results, especially in the main lobes of the beams.  However, we can see differences in the amplitude and shape of the secondary lobes. The main reason for this discrepancy is the effect of the inhomogeneity of induced currents close to the edges of the metasurface, which is neglected in the adopted approximate model. It is important to notice that this discrepancy depends on the size of the metasurface (as expected, for lager metasurfaces, the error decreases and for smaller metasurfaces the error increases). The size of this example ($10D\times 10D$) can be considered as the limiting case for the applicability of this approximate method.

%so that even for the evanescent modes we can assume that the field is nearly uniform over a few unit cells of the microstructure (otherwise the assumption that perturbation of fields near the edges are negligible is not justified). \textcolor{red}{Since in this work we use only the surface impedance model, in all numerical examples we include the contributions of propagating field harmonics only, neglecting the small effects of additional diffuse scattering due to edge inhomogeneities  of the evanescent fields. }

%It is important to notice that,  for a rigorous calculation of the radiation patterns produced by the metasurface, one should calculate the induced currents on the metasurface (considering the effects on the borders) and integrate them to obtain the scattered field. This complicates the calculation of the radiation patterns.  The method proposed in this work provides a simple alternative to visualize and estimate the distribution of power reflected by finite-sized anomalous reflectors.

\section {Angular Bandwidth of Anomalous Reflectors}

Anomalous reflectors are designed to change the direction of reflected waves breaking the reflection law, $\theta_{\rm id}\neq\theta_{\rm rd}$. In a sense, for the design conditions, an anomalous reflector acts as a virtual tilted mirror rotated by the angle $\theta_{\rm t}=(\theta_{\rm id}+\theta_{\rm rd})/2$. Under an arbitrary illumination, a tilted mirror reflects the incident waves to the angle  $\theta_{\rm r}=\theta_{\rm i}+2\theta_{\rm t}$. This linear relation can be understood as the angular dispersion of the ideal tilted mirror.

Because of the functional analogy between the two problems, one can define the angular bandwidth of an anomalous reflector as the range of incident angles within which it behaves as an ideal tilted mirror. As we have learnt from the analysis in Section~\ref{sec:angular}, anomalous reflectors predominantly couple the energy into the first diffracted mode $n=1$. So, in order to analyze the bandwidth of anomalous reflectors we should focus on the angular dispersion of the first diffracted mode. 

\begin{figure}[h]
\center
\subfigure[]{\includegraphics[width=.7\linewidth]{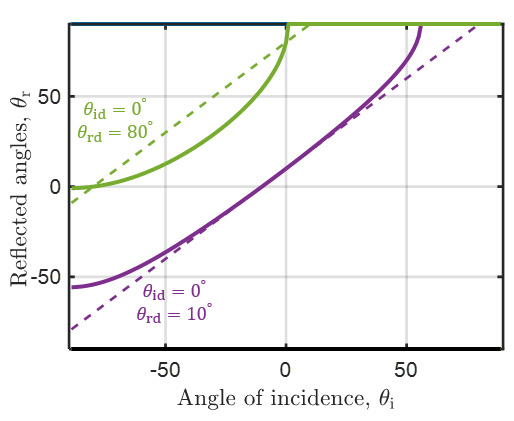}}
\subfigure[]{\includegraphics[width=.45\linewidth]{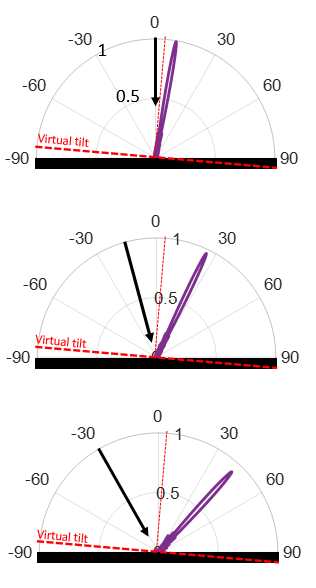}}
\subfigure[]{\includegraphics[width=.45\linewidth]{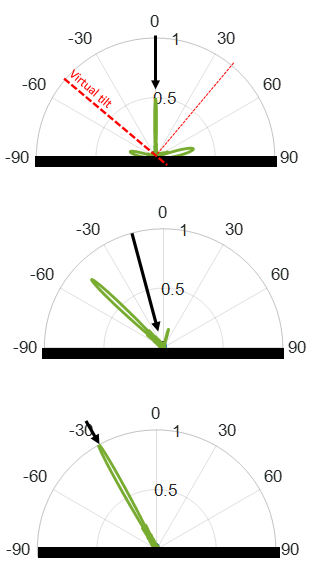}}
\caption{Angular bandwidth of anomalous reflectors. (a) Angular dispersion of the first-order diffracted mode compared with the linear dispersion of a tilted mirror for two different design conditions: $\theta_{\rm id}=0^\circ$ and $\theta_{\rm rd}=10^\circ$ and $\theta_{\rm id}=0^\circ$ and $\theta_{\rm rd}=80^\circ$. (b) Normalized radiation pattern for an anomalous reflector designed for $\theta_{\rm id}=0^\circ$ and $\theta_{\rm rd}=10^\circ$ when $a=b=10\lambda$ and $\theta_{\rm i}=0^\circ, 15^\circ, 30^\circ$. (c) Normalized radiation pattern for an anomalous reflector designed for $\theta_{\rm id}=0^\circ$ and $\theta_{\rm rd}=80^\circ$ when $a=b=10\lambda$ and $\theta_{\rm i}=0^\circ, 15^\circ, 30^\circ$. }
\label{fig:Angular_bandwidth}
\end{figure}

Figure~\ref{fig:Angular_bandwidth}(a) reports the angular dispersion of the first-order diffracted mode compared to the linear dispersion of the equivalent tilted mirror for two different configurations. First, let us consider an anomalous reflector designed to produce a small transformation of the incident waves, $\theta_{\rm id}=0^\circ$ and $\theta_{\rm rd}=10^\circ$. In this case, we can see that the angular dispersion of the diffracted modes can be approximated by the linear dispersion of the tilted mirror when $\theta_{\rm i}\in [-40^\circ, 40^\circ]$. To analyze the scattering properties of this anomalous reflector we will consider a finite-size rectangular metasurface sample ($10\lambda \times 10\lambda$). 

%Using the macroscopic model developed in Section~\ref{sec:rn}, we can write the normalized radiation pattern as
%\begin{eqnarray}
 %   {F}_{{\rm r} z}(\theta)=\frac{1}{2}\sum_nr_n(\theta_{\rm i})(\cos\theta_{{\rm r}n}+\cos\theta){\rm sinc}(ka_{{\rm ef}n})
%\end{eqnarray}
%Note that the radiation pattern is normalized to give unity when $\theta_{\rm i}=\theta_{{\rm r}0}=0$, $r_0=1$, and $r_n=0$ for $n\neq0$. 

Using this definition, the field scattered by the metasurface when  $\theta_{\rm i}=0^\circ, 15^\circ, 30^\circ$ is represented in Fig.~\ref{fig:Angular_bandwidth}(b). From this analysis one can clearly see that, as far as the angular dispersion of the first-order diffracted mode is linear, the metasurface behaves as an efficient virtually tilted mirror. 

Angular dispersion of the diffracted modes strongly depends on the periodicity of the system that is directly related with the desired transformation to be performed by the anomalous reflector. As we can see from the second example shown in  Fig.~\ref{fig:Angular_bandwidth}(a), where $\theta_{\rm id}=0^\circ$ and $\theta_{\rm rd}=80^\circ$, the dispersion of the first-order diffracted mode cannot be approximated by a linear relation. In this example, the anomalous reflector will only behave as a virtually tilted mirror for the design conditions and the reciprocal scenario.  Figure~\ref{fig:Angular_bandwidth}(c) shows the scattering properties of this anomalous reflector when $\theta_{\rm i}=0^\circ, 15^\circ, 30^\circ$ for a square-shaped metasurface of the same size ($10\lambda \times 10 \lambda$). Here, we can see a completely different behaviour than in the previous example. Even a small deviation of the incident angle from the design value leads to a dramatic change in reflector performance, which becomes totally different from the anomalous reflection with the desired tilt by $80^\circ$. This result confirms the strong dependency of the angular bandwidth on the design conditions of  anomalous reflectors.

\section{Influence of the Actual Topology to the Angular Response}
\label{sec:topology}

In the preceding analysis, we assumed that the surface impedance remained unchanged for different illuminations angles. As we mentioned before, this assumption is only valid for some actual structures  and cannot be considered as a general rule. In this section, we present an example of a topology that satisfies this condition and validates this assumption.

\begin{figure}[h]
\center
\subfigure[]{\includegraphics[width=.9\linewidth]{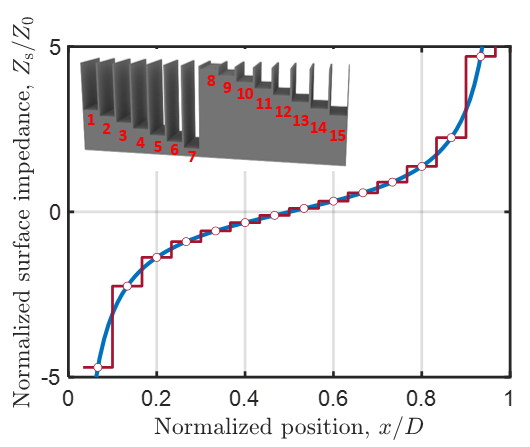}}
\subfigure[]{\includegraphics[width=.9\linewidth]{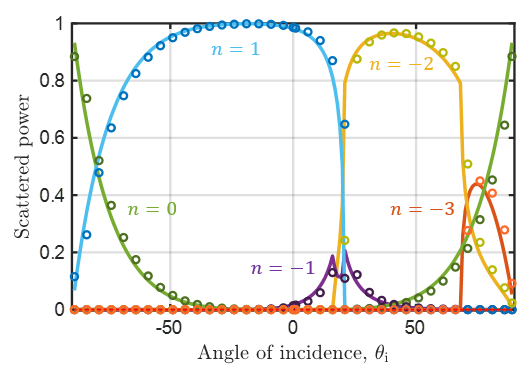}}
\caption{Analysis of the scattering properties of an infinite anomalous reflector designed for $\theta_{\rm id}=0^\circ$ and $\theta_{\rm rd}=40^\circ$ and implemented using metallic grooves. (a) Normalized surface impedance for the anomalous reflector. Solid line represents the continuous surface impedance dictated by Eq.~\ref{eq:surface_impedance}. Line with symbols represent the surface impedance implemented by the metallic grooves. Inset picture represent the scheme of the structure that implements the anomalous reflector. (b) Energy distribution between the propagating harmonics for different incident angles. Solid lines represent the results considering a continuous surface impedance. Symbols represent the results obtained from the numerical simulation of the actual structure.}
\label{fig:Conventiona_0_40_infinite}
\end{figure}

In particular, we will numerically confirm that  one can implement a surface impedance dictated by Eq.~\eqref{eq:surface_impedance} without angular dispersion using closed-end grooves in a metal plate [see in Fig.~\ref{fig:Conventiona_0_40_infinite}(a)]. These  meta-atoms have been used to implement power flow-conformal metasurfaces \cite{diaz2020dual}. We assume that the grooves are infinitely long and uniform and the thickness of the walls is much  smaller than the width of the grooves, so that we can neglect the effects of wall thickness on the surface impedance. For TM-polarized electromagnetic waves, the response of the meta-atom is defined  by the surface impedance $Z_{\rm s}^{\rm EM}=jZ_0\tan(k\ell)$, with $\ell$ being the depth of the groove. As an example, we will design and numerically study an anomalous reflector targeted  for  $\theta_{\rm id}=0^\circ$ and $\theta_{\rm rd}=40^\circ$. The operation frequency of the metasurface is $f=8$~GHz. The surface impedance required to implement this functionality is represented in Fig.~\ref{fig:Conventiona_0_40_infinite}(a) with a solid line. To have a fine surface profile and avoid errors due to the discretization of the continuous surface impedance, we use 15 elements per period to  implement the surface impedance. The depths of the grooves to produce the required surface impedance are 10.625~mm, 11.875~mm, 13.125~mm, 14.375~mm, 15.625~mm, 16.875~mm, 18.125~mm, 0.625~mm, 1.875~mm, 3.125~mm, 4.375~mm, 5.625~mm, 6.875~mm, 8.125~mm, and 9.375~mm. The implemented surface impedance is represented in Fig.~\ref{fig:Conventiona_0_40_infinite}(b) with a dashed line.

To make the metasurface thickness small compared to the free-space wavelength, the grooves can be filled with a high-permittivity dielectric. In this case, each unit cell  can be be viewed as a combination of an electric-current element (due to the current at the bottom termination) and a magnetic-current element (due to the out-of-phase currents at the vertical walls). Effective induced magnetic current is absent only in the case of zero depth of the groove. As was discussed in Section~\ref{sec:rn}, this  obviously corresponds to the reflection phase $180^\circ$.

The numerical study of the structure is done using the commercial software Ansys HFSS. The response of the structure is represented in Fig.~\ref{fig:Conventiona_0_40_infinite}(c) by the circular symbols. For comparison, the angular response for a metasurface implementing the continuous surface impedance and calculated using the Floquet harmonic expansion has been included in the figure (see solid lines). We can see  excellent agreement between both results verifying the angular independence of the implemented surface profile. We note that for other realizations of phase-gradient reflectors (most commonly, as planar patch arrays at a grounded dielectric substrate when the substrate permittivity is not large) the surface impedance depends on the incident angle, which makes the angular dependence of the metasurface response more complicated.

\section {Conclusions}

In this work we have investigated the angular response of phase-gradient metasurfaces designed for manipulating the direction of propagation of reflected electromagnetic waves. Due to their extraordinary properties for wave manipulation and possibilities for simple integration in real-world environments,  characterizing the scattering properties of these metasurfaces under different illumination conditions is necessary for the development of smart radio environments and other applications of advanced metasurfaces.

Using the surface impedance model, we have studied the response of anomalous reflectors for arbitrary illumination angles. The results show that for anomalous reflectors designed using the linear phase gradient of the local reflection coefficient, the incident power is predominantly coupled to the diffracted mode $n=1$. Only when the design requirements demand strong transformations of the direction of propagation, significant portion of the power is reflected specularly. We have also delved into the relation of the number of diffracted modes and the angular response for reciprocal metasurfaces. We have verified the results obtained with the surface impedance model by designing an actual structure that confirms the angular response predicted by the impedance model.

We have proposed an analytical  modeling approach to predict the far-zone reflective and scattering responses of finite-size anomalous reflectors for arbitrary illumination angles. This model can be used as the key element in future analytical and numerical models of radio wave propagation in complex environments including metasurface panels on the surfaces of walls or facades. 

To find the reflected fields in the far zone, we merge some ideas of the physical optics (but using macroscopic, essentially nonlocal expressions for the induced currents) and the approaches of the theory of diffraction by gratings (because anomalous reflectors are periodical structures that can create multiple diffracted beams). 
The developed method is based on the use of introduced macroscopic reflection coefficients that relate the amplitude of the incident wave and the amplitudes and phases of the fields of each diffracted mode. Using the developed formalism, it is simple to calculate the far-field response of the metasurface when it is illuminated from any direction and estimate the link budget for arbitrary positions of the transmitter and receiver. 

The developed analytical formulas for far-field scattering from finite-size metasurfaces are valid not only for phase-gradient reflectors but also for nonlocal metasurfaces that offer theoretically perfect anomalous reflection. In those advanced designs \cite{Diaz_From_2017,Epstein_Unveiling_2017,Radi_Metagrating_2018,epstein2016synthesis,Diaz_Power_2019}, the amplitude of reflection from the design direction $\theta_{\rm i}=\theta_{\rm id}$ into the desired anomalously reflected mode is close to the ideal value defined by Eqs.~\eqref{eff_TE} and \eqref{eff_TM} for 100\% efficiency. 
%$|r_1|=\sqrt{\cos\theta_{\rm i}\over \cos\theta_{\rm rd}}$ (TE polarization) 
The amplitudes of all other propagating Floquet harmonics are suppressed to negligible levels. In this case, practically all incident power is reflected as a single plane wave propagating in the desired direction, if the metasurface is infinite. For other angles of incidence, there is scatterings into all propagating modes, and the corresponding amplitudes need to be found numerically or experimentally. 
Importantly, the phase of the main reflected harmonic can be engineered, which opens up a possibility to control the side-lobe structure of scattering from finite-size metasurfaces because scattering into side lobes depends on interference with the shadow radiation, as illustrated in Fig.~\ref{fig:Radiation_diagram}.

\section*{Acknowledgment}

This work was supported in part by the European Commission through the H2020 ARIADNE project under grant 871464 and the Academy of Finland under the grant 330957.

\bibliographystyle{IEEEtran}
\bibliography{references}

%\begin{thebibliography}{00}

%\bibitem{b1} ....

%\end{thebibliography}

\end{document}